\documentclass[11pt,notitlepage,tightenlines,eqsecnum,floats,aps,amsmath,superscriptaddress,amssymb,nofootinbib,longbibliography]{revtex4-1}

\usepackage{graphicx}
\usepackage{amsmath,amssymb}
\usepackage{hyperref}
\usepackage[normalem]{ulem}
\usepackage{comment}
\usepackage{wrapfig}
\usepackage{braket}

\begin{document}

\title{Effective Regge-Wheeler equations of a hybrid loop quantum black hole}

\author{Beatriz Elizaga Navascués}
\email{beatriz.elizaga@uclm.es}
\affiliation{Department of Physics, Faculty of Environmental Sciences and Biochemistry, University of Castilla - La Mancha, 45071 Toledo, Spain.
 }
 \author{Alvaro Torres-Caballeros}
\email{alvaro.torres@iem.cfmac.csic.es}
\affiliation{Instituto de Estructura de la Materia, IEM-CSIC, C/ Serrano 121, 28006 Madrid, Spain.}

\begin{abstract}
A set of effective equations for the gauge-invariant gravitational perturbations in the interior of a spherically symmetric, non-rotating black hole is derived within the framework of hybrid loop quantum cosmology. The quantum zero-mode of the Hamiltonian constraint, obtained from a perturbative gauge-invariant canonical analysis, is explicitly imposed on a class of quantum states whose wavefunctions factorize into a background and a perturbative part, related through a geometric relational variable. These states naturally describe regimes with small perturbative backreaction and lead to an effective Hamiltonian and associated dynamics for the perturbations. The resulting equations take the form of Regge–Wheeler equations modified by expectation values of the quantum black hole geometry, providing a clear characterization of quantum corrections to the classical description of the black hole interior. This framework opens the way to investigating hybrid loop quantum gravity effects in the propagation of gravitational waves.
\end{abstract}

\maketitle

\section{\label{sec:1} Introduction}

The advent of gravitational wave astronomy has opened a new way of testing the classical theory of General Relativity (GR) \cite{LIGOScientific:2020tif,LIGOScientific:2021sio}, in the strong-field regime. The possibility that future observations in this area may reveal departures from (or tensions with) GR is an exciting prospect, from a theoretical perspective. In particular, it should not be ruled out {\emph {a priori}} that gravitational waves of astrophysical origin may carry, in some way, information about the quantum nature of gravity. In this context, the importance of theoretical approaches to the quantization of GR lies in the ability to provide concrete predictions about departures from the classical theory (if any), expected to be observed in astrophysical scenarios.

So far, most of the gravitational wave sources that have been directly detected are black hole binaries. Their coalescence is described by a non-linear process in GR, the full analysis of which is highly complicated and heavily relies on numerical tools \cite{LISAConsortiumWaveformWorkingGroup:2023arg}. In the current state of the field, it would be most impractical to try to describe this whole process within a quantum gravity framework. There is, however, a regime that is amenable to a semi-analytical and linear description: The so-called ``ringdown", which is the stabilization process of the post-merger black hole \cite{Berti:2025hly}. In GR (and in the absence of electric charge), this process can be described using perturbation theory around an exact solution of Schwarzschild \cite{Regge:1957td,Zerilli:1970se,Moncrief:1974am} or Kerr type \cite{Teukolsky:1973ha,Press:1973zz}. The gauge-invariant perturbations of these background spacetimes correspond (in vacuo) to the gravitational waves emitted in the last stages of the coalescence. The comparative simplicity of this perturbative system is a promising feature that encourages developing a full quantum description of it.

Among the different proposals to achieve a quantization of the gravitational interaction, over the last two decades, the formalism of Loop Quantum Gravity (LQG) \cite{Rovelli:2004tv,Thiemann:2007pyv,Ashtekar:2021kfp} has stood out regarding predictive power in scenarios of physical interest. In particular, its application to the quantization of cosmological spacetimes, known as Loop Quantum Cosmology (LQC) \cite{Ashtekar:2011ni}, has grown to become a mature field able to make concrete predictions on possible observational effects of quantum gravity on the Cosmic Microwave Background (see e.g. Refs.~\cite{Ashtekar:2020gec,ElizagaNavascues:2020uyf,Agullo:2023rqq} and references therein). The physical phenomenon behind these effects is the replacement of the Big-Bang singularity by a bounce of quantum origin, a renowned result of LQC \cite{Ashtekar:2006rx,Ashtekar:2006wn}. It is only natural to ask whether a similar resolution of the singularity in the context of black holes may somehow result in departures from GR in their gravitational wave emission.

The interior of spherically symmetric black holes in vacuum GR is isometric to a cosmological spacetime of Kantowski-Sachs type \cite{Kantowski:1966te}. This observation has motivated several investigations of its description within the formalism of LQC. These works range from genuine quantizations of the interior geometry (see e.g. Refs.~\cite{Zhang:2020qxw,Zhang:2021wex,ElizagaNavascues:2022rof,ElizagaNavascues:2023gow} and references therein), to {\emph {effective}} descriptions of the spacetime metric that one would expect to arise from states that are peaked on geometric observables (see e.g. Refs.~\cite{Olmedo:2017lvt,Cortez:2017alh,Ashtekar:2018cay} and references therein). Remarkably, all of them point towards a resolution of the interior singularity \cite{Ashtekar:2023cod}. Moreover, the physical Hilbert space of the system has been completely characterized recently, revealing in two possible situations: That the black hole mass be discrete \cite{Zhang:2021wex}, or that it can take any continuous value \cite{ElizagaNavascues:2023gow}.
%\cite{Ashtekar:2005qt,Modesto:2005zm,Cartin:2006yv,Boehmer:2007ket,Campiglia:2007pr,Corichi:2015xia,Yonika:2017qgo,Zhang:2020qxw,Zhang:2021wex,ElizagaNavascues:2022rof,ElizagaNavascues:2023gow}
%\cite{Chiou:2008nm,Modesto:2008im,Olmedo:2017lvt,Cortez:2017alh,Ashtekar:2018lag,Ashtekar:2018cay,Ashtekar:2020ckv,ElizagaNavascues:2022npm}.

Having reached sufficient control over the quantum description of the non-rotating black hole interior in LQC, recent works have taken over the task of introducing gravitational perturbations in the system \cite{MenaMarugan:2024qnj,MenaMarugan:2025anx}. These investigations start from the Hamiltonian formulation of perturbation theory around spherically symmetric spacetimes \cite{Brizuela:2008sk}, and apply a hybrid quantization to the system truncated at the lowest non-trivial perturbative order. In a nutshell, this hybrid quantization scheme consists of combining some quantum gravity representation (e.g. LQC) for the spatially homogeneous sector of the system, namely the background spacetime, with a suitable Fock representation for the inhomogeneous perturbations. It was first introduced in the context of Gowdy cosmologies \cite{Martin-Benito:2008eza}, and it reached peak success in the treatment of primordial fluctuations in LQC \cite{Gomar:2015oea,ElizagaNavascues:2020uyf}, opening the way to phenomenological studies.

Remarkably, the description attained in Refs.~\cite{MenaMarugan:2024qnj,MenaMarugan:2025anx} isolates the degrees of freedom of the gravitational field that are perturbatively gauge-invariant, at the considered (lowest) order \cite{Brizuela:2006ne,Brizuela:2007zza}. The resulting Hamiltonian is a linear combination of constraints, containing: i)~The purely homogeneous contribution (or zero-mode) of the Hamiltonian constraint of GR, truncated at quadratic perturbative order; and ii)~A set of linear perturbative constraints, which are canonically conjugated to unphysical, pure gauge, degrees of freedom. The latter constraints are trivial to impose in the quantum theory: They simply restrict physical states not to depend on pure gauge. It is the zero-mode Hamiltonian constraint, depending only on the background geometry and the gauge invariants, the object that contains all the relevant physics of the system. The authors of Refs.~\cite{MenaMarugan:2024qnj,MenaMarugan:2025anx} discuss how this constraint should be quantized, but do not explore the specific consequences of such quantization on the physical states of the system.

The goal of this work is two-fold, starting from the hybrid loop quantum black hole interior formulated in Refs.~\cite{MenaMarugan:2024qnj,MenaMarugan:2025anx} (and recently refined in Ref.~\cite{Lenzi:2025pmk}). First, we explicitly impose the quantum Hamiltonian constraint, discussing in detail what type of candidate states are suitable to describe scenarios of physical interest.  Second, we derive an effective Hamiltonian, and its associated dynamics, for the gauge-invariant perturbations. The result is a set of effective equations for gravitational waves in the interior, where one can clearly identify the modifications due to LQC with respect to their classical counterpart in black hole spacetimes: The Regge-Wheeler equations. 

Our work can be viewed as a necessary adaptation, to a black hole setting, of the steps leading to effective equations for gauge invariants in the context of primordial LQC \cite{Gomar:2015oea}. It presents, nonetheless, an important novelty with respect to the cosmological case: The effective dynamics is derived without the need of introducing a matter field that defines a time coordinate for the system, but by using the geometry itself as a relational clock. As a consequence, our work includes an analysis of which variable describing the black hole geometry is suitable to define a relational dynamics in the quantum theory. For concreteness in such analysis, we shall focus on those situations where the loop quantization allows the mass of the black hole to take any continuous value. This strategy also frees us from dealing with the possible consequences of a mass gap in radiative phenomena, such as the Hawking effect or the presence of gravitational wave echoes \cite{Agullo:2020hxe}. We leave the study of scenarios with discrete mass, and their comparison with present results, for the future.

The article is organized as follows. In Sec.~\ref{sec:2} we summarize the model under study, namely a perturbed cosmological spacetime describing a black hole interior, and its hybrid quantization in terms of gauge invariants. We proceed in Sec.~\ref{sec:3} to discuss what type of quantum states would best describe a scenario where the backreaction of the perturbations on the background is small, considering the properties of the physical Hilbert space of the unperturbed model in LQC. We then impose and work out the action of the Hamiltonian constraint operator on these states. The resulting constraint equation allows us, in Sec.~\ref{sec:4}, to identify an effective Hamiltonian for the perturbations and, with a suitable definition of time, derive a set of Regge-Wheeler equations with LQC modifications. Finally, we conclude and provide an outlook for our work in Sec.~\ref{sec:5}. We use Planck units throughout the article, setting $G=\hbar=c=1$. Moreover, bars and hats over quantities respectively denote complex conjugation and quantum representation as operators on a Hilbert space.

\section{\label{sec:2} The classical model and its quantization}

Let us consider a spacetime manifold that can be foliated in spatial slices with the symmetries of a Kantowski-Sachs cosmology, up to the addition of inhomogeneous metric perturbations. This spacetime model is the starting point for our description of the interior of a spherically symmetric black hole with perturbations of gravitational type. In the following subsections, we provide all the relevant details of the model and its hybrid quantization. 

For the sake of clarity, we first discuss the quantization of spacetime without perturbations within the framework of LQC. This homogeneous cosmological model constitutes the exact background sector of the phase space of our model, within the perturbative scheme. Then, we elaborate on how the gauge-invariant perturbations contribute to the Hamiltonian of the system, when it is truncated at quadratic order, and on the hybrid quantization of such Hamiltonian.

It should be noted that this section is but a necessary summary of previous work. We refer the reader to Refs.~\cite{Zhang:2021wex,ElizagaNavascues:2023gow,ElizagaNavascues:2022rof} for a detailed introduction to the homogeneous background model in LQC, and to Refs.~\cite{MenaMarugan:2024qnj,MenaMarugan:2025anx,Lenzi:2025pmk} for the Hamiltonian treatment and quantization of the system with gauge-invariant perturbations.

\subsection{Homogeneous background}\label{subsec:2.1}

A spacetime with the symmetries of Kantowski-Sachs can be identified with a homogeneous but anisotropic cosmology with spatial slices that are homeomorphic to $\mathbb{R}\times S^2$, where $S^d$ denotes the $d-$dimensional sphere. It is also possible to fully compactify these slices and work with $S^1\times S^2$. For mathematical convenience, let us focus on this latter possibility and denote by $L_o$ the compactification period of $S^1$. The non-compact case can be recovered by taking the limit corresponding to $L_o\to \infty$, in an appropriate manner.

The line element of the considered type of spacetime can be written as follows:
\begin{align}\label{backgroundmetric}
ds^2=-\frac{p_b^2|p_c|}{L_o^2}N_0^2\,d\tau^2+\frac{p_b^2}{L_o^2|p_c|}\,dx^2+|p_c|\,d\Omega^2_2~,
\end{align}
where $p_b$ and $p_c$ are two spatially homogeneous functions, $d\Omega_2^2$ is the line element of $S^2$, and $N_0$ is the lapse function of the $3+1$ foliation conveniently multiplied by $4\pi  L_o/\mathcal{V}$, where $\mathcal{V}=4\pi |p_b|\sqrt{|p_c|}$ is the physical volume of the spatial slices. With our choice of topology for these slices, we note that the coordinate $x$ takes values in the interval $[0,L_o)$. Moreover, the quantities $p_b/L_o$ and $p_c$ can be seen to be invariant under rescalings of the compactification length $L_o$ \cite{Ashtekar:2018cay}.

The variables $p_b$ and $p_c$ play a canonical role in a Hamiltonian formulation of the system. Specifically, they are canonically conjugated to another pair of spatially homogeneous functions $(b,c)$ which can be chosen in such a way that
\begin{align}
\{b,p_b\}=\gamma~,\qquad \{c,p_c\}=2\gamma~,
\end{align}
with all other Poisson brackets being identically zero, and where $\gamma$ is the Immirzi parameter in LQG. The pairs $(b,c)$ and $(p_b,p_c)$ respectively characterize the Ashtekar-Barbero connection and densitized triad of the cosmological model \cite{Ashtekar:2018cay,ElizagaNavascues:2022rof}. Their use as canonical variables is, thus, most convenient in view of the loop quantization of the system.

The gravitational Hamiltonian of the considered spacetime, with a $3+1$ foliation of the type of Eq.~\eqref{backgroundmetric}, is given by $H_{\rm{KS}}[N_0]=N_0\,C_0$, where $C_0$ is the following function on phase space:
\begin{align}\label{KSH}
C_0=-\frac{L_o}{2}\left(\Omega^2_b+\frac{p_b^2}{L_o^2}+2\Omega_b\Omega_c\right)~,
\end{align}
and $\Omega_j=j\,p_j/(\gamma L_o)$ for $j=b,c$. This function vanishes on solutions, as a consequence of applying the variational principle with respect to the Lagrange multiplier $N_0$. Moreover, $\Omega_c$ is clearly a constant of motion, which coincides with the ADM mass of the black hole (up to sign) when the corresponding solutions are identified with the interior region of a Kruskal spacetime \cite{Ashtekar:2018cay}.

At the kinematic level, LQC provides a non-continuous quantum representation of the Poisson algebra formed by the triadic pair $(p_b,p_c)$ and holonomies of the Ashtekar-Barbero connection along spatial edges in the directions of symmetry. As a result, there are no well-defined operators representing the pair $(b,c)$ on the kinematic Hilbert space of the theory. This poses an obstruction to the definition of a (densitized) Hamiltonian constraint operator, due to the presence of the variables $\Omega_j$ in Eq.~\eqref{KSH}, originating from contributions of the curvature of the connection to the Hamiltonian constraint in GR \cite{ElizagaNavascues:2022rof}. This issue is circumvented in LQC by postulating the existence of a minimum non-vanishing value $\Delta$ of the physical area of spatial plaquettes in the system, and replacing the curvature of the connection by circuits of holonomies along the edges of such plaquettes \cite{Ashtekar:2018cay,ElizagaNavascues:2022rof}. This procedure results in a redefinition of the variables $\Omega_j$ appearing in the Hamiltonian into the following function on phase space: $\Omega_j=\sin(\delta_j\,j)\,p_j/(\delta_j\gamma L_o)$, where the parameters $\delta_j$ are directly related to the minimum value allowed for the area.

Several different prescriptions have been proposed to fix the parameters $\delta_j$ in the LQC literature. Remarkably, effective spacetimes exhibiting several physically desirable properties can be derived if one constrains $\delta_j$ to be related to $\Delta$ in a way that depends on the mass of the black hole \cite{Cortez:2017alh,Ashtekar:2018cay,Olmedo:2017lvt}. These properties include the presence of a regular black-to-white hole transition and the recovery of Schwarzschild spacetime at low curvatures. All results achieved in this work are valid for any such prescription: In the following, we assume that the relation between $\delta_j$ and $\Delta$ is exclusively mediated by the black hole mass, but do not specify such a relation.

Classically, the mass arises as the constant of motion $\Omega_c$, making it natural to identify it with this quantity in the quantum theory as well. With this identification, the redefined functions $\Omega_j=\sin(\delta_j\,j)\,p_j/(\delta_j\gamma L_o)$ become amenable to quantization on the kinematic Hilbert space of LQC, when the parameters $\delta_j$ are constrained to be functions of the black hole mass and $\Delta$ \cite{Zhang:2021wex,ElizagaNavascues:2023gow}. The resulting operators $\hat{\Omega}_j$ are essentially self-adjoint and have an absolutely continuous spectrum equal to $\mathbb{R}$. Moreover, they superselect the kinematic Hilbert space into separable sectors, allowing one to restrict all subsequent analysis to any one of them. Explicitly, if we symbolically call $|\mu_b,\mu_c\rangle$ the elements of the orthonormal basis of any separable sector $\mathcal{H}_b\otimes\mathcal{H}_c$, defined so that
\begin{align}\label{polymeric}
\frac{\hat{p}_b}{\delta_b}|\mu_b\rangle = \frac{\gamma\mu_b}{2}|\mu_b\rangle~,\quad \frac{\hat{p}_c}{\delta_c}|\mu_c\rangle = \gamma\mu_c \, |\mu_c\rangle~,\qquad \langle \mu'_j|\mu_j\rangle=\delta_{\mu'_j,\mu_j}~,\quad j=b,c
\end{align}
where $\delta_{y',y}$ is the Kronecker delta, then the labels $\mu_j$ are restricted to take values in any fixed set of the form $\{\pm(\epsilon_j+2n),\,n\in\mathbb{N}\}$, with $\epsilon_j\in(0,2]$ \cite{ElizagaNavascues:2023gow}.

For future use, it is convenient to remark that the spectral properties of $\hat{\Omega}_j$ allow one to identify wave functions $\breve\psi(\mu_j)$ in $\mathcal{H}_j$ with their generalized Fourier transform in $L^2(\mathbb{R})$:
\begin{align}\label{gfourier}
\psi(\omega_j)=\sum_{\mu_j}\breve\psi(\mu_j)\,\bar{e}^{\epsilon_j}_{\omega_j}(\mu_j)~,
\end{align}
where $e^{\epsilon_j}_{\omega_j}$ denotes the eigenfunction of $\hat{\Omega}_j$ with eigenvalue $\omega_j$, normalized to the Dirac delta:
\begin{align}
\sum_{\mu_j}\bar{e}^{\epsilon_j}_{\omega'_j}(\mu_j)\,e^{\epsilon_j}_{\omega_j}(\mu_j)=\delta(\omega_j'-\omega_j)~.
\end{align}
Finally, we notice that the quantum number $\omega_c$ labels the black hole mass in the quantum theory.
 
\subsection{Perturbed system}

We proceed to consider inhomogeneous fluctuations around the cosmological spacetime introduced above, which we identify as the exact background spacetime in the perturbative scheme. Specifically, calling $\mathbf{N}$, $N^{\alpha}$, and $h_{\alpha\beta}$ the lapse function, the shift vector and the spatial metric of the perturbed spacetime, we consider any foliation such that we can write
\begin{equation}
\begin{split}
\mathbf{N}=\frac{|p_b|}{L_o}\sqrt{|p_c|}\left(N_0+\delta N\right)~,\quad N_{\alpha}\,dx^{\alpha}=\delta N_{\alpha}\,dx^{\alpha}~, \\ h_{\alpha\beta}\,dx^{\alpha}dx^{\beta}=\frac{p_b^2}{L_o^2|p_c|}\,dx^2+|p_c|\,d\Omega^2_2+\delta h_{\alpha\beta}\,dx^{\alpha}dx^{\beta}~,
\end{split}
\end{equation}
where $\delta N$, $\delta N_{\alpha}$, $\delta h_{\alpha\beta}$  denote purely inhomogeneous fields to be treated as perturbations.

Given the symmetries of the homogeneous background, it proves most convenient to decompose the metric perturbations into Fourier modes and the so-called Regge-Wheeler-Zerilli harmonics. These respectively provide bases for the expansion of tensor fields on $S^1$ and $S^2$. We adopt the following notation for them:
\begin{enumerate}
    \item[1)] Real Fourier modes on $S^1$ are denoted by $Q_{n,\lambda}$, with $\lambda=\pm$ indicating their parity under the transformation $x\rightarrow -x$, and $n\in\mathbb{N}$.
    \item[2)] Real spherical harmonics on $S^2$ are denoted by $Y^{m}_l$, with eigenvalue of the Laplacian on $S^2$ equal to $-l (l+1)$, where $l\in\mathbb{N}$ and $m=-l,-l+1,\ldots, l-1,\, l$. They have parity $(-1)^{l}$ under point inversions on the sphere.
    \item[3)] Covector harmonics on $S^2$ are denoted by $X_{l}{}^{m}{}_A$ and $Z_{l}{}^{m}{}_A$, with $l\in\mathbb{N}-\{0\}$. They have respective parity $(-1)^{l+1}$  and $(-1)^{l}$ under point inversions on the sphere.
    \item[4)] Tensor harmonics on $S^2$ are denoted by $X_l{}^{m}{}_{AB}$ and $Z_l{}^{m}{}_{AB}$, with $l\in\mathbb{N}-\{0,1\}$. These harmonics are traceless and have respective parity $(-1)^{l+1}$ and $(-1)^{l}$ under point inversions on the sphere. Together with the trace harmonic $Y^{m}_{l}\gamma_{AB}$, where $\gamma_{AB}$ denotes the metric on $S^2$, they form a complete basis for the expansion of symmetric tensor fields of rank $2$ on the sphere. 
\end{enumerate} 
We refer the reader to Refs.~\cite{Brizuela:2006ne,Brizuela:2007zza} for a more detailed description of the Regge-Wheeler-Zerilli harmonics used in this work. Explicitly, we employ the following type of decomposition \cite{MenaMarugan:2024qnj,MenaMarugan:2025anx,Lenzi:2025pmk}:
\begin{equation}
\begin{split}
\delta N&=-\frac{1}{2}N_0\sum_{\lambda\in\{\pm\}}\sum_{\mathfrak{n}\in\mathfrak{N}_0}f_0^{\mathfrak{n},\lambda} Y^{m}_{l}Q_{n,\lambda}~, \\ \delta N_{\alpha}\,dx^{\alpha}&= \sum_{\lambda\in\{\pm\}}\left[\frac{p_b^2}{L_o^2}\sum_{\mathfrak{n}\in\mathfrak{N}_0}k_0^{\mathfrak{n},\lambda}Y^{m}_{l}Q_{n,\lambda}\,dx +\sum_{\mathfrak{n}\in\mathfrak{N}_1}\left(|p_c|q_{0}^{\mathfrak{n},\lambda}Z_l{}^{m}{}_{A}-h_0^{\mathfrak{n},\lambda}X_l{}^{m}{}_{A}\right)Q_{n,\lambda}\,dx^{A}\right]~
\end{split}
\end{equation}
for the perturbations of the lapse and shift vector, where $x^{A}$ denotes the angular coordinates on $S^2$, whereas
\begin{equation}
\begin{split}
\delta h_{\alpha\beta}\,dx^{\alpha}dx^{\beta}=\sum_{\lambda\in\{\pm\}}\Bigg[&\sum_{\mathfrak{n}\in\mathfrak{N}_0}\left(h_{6}^{\mathfrak{n},\lambda}Y^{m}_{l}\,dx^2+h_3^{\mathfrak{n},\lambda}Y_{l}^{m}{\gamma}_{AB}\,dx^{A}dx^{B}\right)Q_{n,\lambda}\\&+2\sum_{\mathfrak{n}\in\mathfrak{N}_1}\left(h_5^{\mathfrak{n},\lambda}Z_{l}{}^{m}{}_{A}-h_{1}^{\mathfrak{n},\lambda}X_{l}{}^{m}{}_A\right)Q_{n,\lambda}\,dxdx^{A}\\&+\sum_{\mathfrak{n}\in\mathfrak{N}_2}\left(h_2^{\mathfrak{n},\lambda}X_{l}{}^{m}{}_{AB}+h_{4}^{\mathfrak{n},\lambda}Z_{l}{}^{m}{}_{AB}\right)Q_{n,\lambda}\,dx^Adx^{B}\Bigg]
\end{split}
\end{equation}
is the decomposition for the metric perturbations. In these formulas, the modes $f_0^{\mathfrak{n},\lambda},k_0^{\mathfrak{n},\lambda},q_{0}^{\mathfrak{n},\lambda},h_0^{\mathfrak{n},\lambda}$ and $h_{i}^{\mathfrak{n},\lambda}$, with $i=1,\ldots,6$, are spatially homogeneous functions. They are labeled with the tuple $\mathfrak{n}=(n,l,m)$, which can take values in sets of the type
\begin{align}
\mathfrak{N}_k=\{(n,l,m)\neq (0,0,0)\,|\,n\in\mathbb{N},l\in\{k,k+1,k+2,\ldots\},m\in\{-l,\ldots,l\}\}~,
\end{align}
with $k=0,1,2$.

Metric perturbations of spherically symmetric spacetimes are sorted into axial and polar types, respectively, depending on whether they have parity $(-1)^{l+1}$ or $(-1)^{l}$  under point inversions on the sphere. In our harmonic decomposition, this means that the modes $h_0^{\mathfrak{n},\lambda},h_1^{\mathfrak{n},\lambda},$ and $h_2^{\mathfrak{n},\lambda}$ describe axial perturbations, whereas $f_0^{\mathfrak{n},\lambda},k_0^{\mathfrak{n},\lambda},q_{0}^{\mathfrak{n},\lambda},h_3^{\mathfrak{n},\lambda},h_4^{\mathfrak{n},\lambda},h_5^{\mathfrak{n},\lambda},$ and $h_6^{\mathfrak{n},\lambda}$ describe polar perturbations.

The total gravitational Hamiltonian of the perturbed system, when truncated at quadratic perturbative order, is a linear combination of constraints with Lagrange multipliers given by $N_0$ and the modes $f_0^{\mathfrak{n},\lambda},k_0^{\mathfrak{n},\lambda},q_{0}^{\mathfrak{n},\lambda},$ and $h_0^{\mathfrak{n},\lambda}$, which therefore describe non-physical degrees of freedom. The constraints can be separated into two sets. 

On the one hand, $N_0$ appears multiplying the zero-mode of the Hamiltonian constraint. It is the sum of the constraint $C_0$ of the unperturbed system, given in Eq.~\ref{KSH},
and a term that is quadratic in the perturbative modes $h_{i}^{\mathfrak{n},\lambda}$, $i=1,\ldots,6$, and their canonical momenta. 

On the other hand, we find the linearization of the Hamiltonian and diffeomorphism constraints of GR, which appear at quadratic order in the action multiplied by the perturbations of the lapse and shift. 

The truncation of the action at quadratic order in perturbations, with its associated symplectic structure and Hamiltonian, is the fully constrained system we want to quantize. For this purpose, and given the freedom in changing the spacetime foliation with perturbative diffeomorphisms, a preliminary step is to identify which variables describe the physical part of the spatially inhomogeneous degrees of freedom.

Conveniently, the authors of Refs.~\cite{MenaMarugan:2024qnj,MenaMarugan:2025anx,Lenzi:2025pmk,Lenzi:2025man} introduced a canonical transformation on the phase space of the full system (background plus perturbations), which is exact at quadratic order in the action and disentangles all dependence of the Hamiltonian on perturbative gauge-invariant quantities and pure gauge degrees of freedom. More specifically, let us denote this canonical transformation as follows:
\begin{align}
\left(b,p_b,c,p_c,\{h_{i}^{\mathfrak{n},\lambda},p_{i}^{\mathfrak{n},\lambda}\}_{i=1,...,6}\right)\longrightarrow \left(\tilde{b},\tilde{p_b},\tilde{c},\tilde{p_c},\{Q_{i}^{\mathfrak{n},\lambda},P_{i}^{\mathfrak{n},\lambda}\}_{i=1,...,6}\right)~,
\end{align}
where $p_{i}^{\mathfrak{n},\lambda}$ is the canonically conjugated momentum to the mode $h_{i}^{\mathfrak{n},\lambda}$, with $\{h_{i}^{\mathfrak{n},\lambda},p_{i}^{\mathfrak{n},\lambda}\}=16\pi$. Our notation is such that the pairs $(Q_{i}^{\mathfrak{n},\lambda},P_{i}^{\mathfrak{n},\lambda})$ are axial for $i=1,2$, and polar otherwise. In terms of these new variables, the Hamiltonian of the system reads as follows:
\begin{equation}\label{GIHam}
\begin{split}
\mathbf{H}=&\tilde{N}_0\left(\tilde{C}_0+\Theta^{\rm{ax}}+\Theta^{\rm{po}}\right)\\&+\frac{1}{16\pi}\sum_{\lambda\in\{\pm\}}\left(\sum_{\mathfrak{n}\in\mathfrak{N}_1}\tilde{h}_0^{\mathfrak{n},\lambda}P_{2}^{\mathfrak{n},\lambda}+\sum_{\mathfrak{n}\in\mathfrak{N}_0}\tilde{f}_0^{\mathfrak{n},\lambda}P_4^{\mathfrak{n},\lambda}+\sum_{\mathfrak{n}\in\mathfrak{N}_1}\tilde{q}_0^{\mathfrak{n},\lambda}P_5^{\mathfrak{n},\lambda}+\sum_{\mathfrak{n}\in\mathfrak{N}_0}\tilde{k}_0^{\mathfrak{n},\lambda}P_6^{\mathfrak{n},\lambda}\right)~.
\end{split}
\end{equation}
In this expression, $\Theta^{\rm{ax}}$ and $\Theta^{\rm{po}}$ are spatially homogeneous functions that depend quadratically on the respective pairs $(Q_{1}^{\mathfrak{n},\lambda}, P_{1}^{\mathfrak{n},\lambda})$ and $(Q_{3}^{\mathfrak{n},\lambda}, P_{3}^{\mathfrak{n},\lambda})$. Moreover, $\tilde{C}_0$ is the constraint of the unperturbed model, given in Eq.~\eqref{KSH}, directly evaluated on the new background variables. Finally, the new Lagrange multipliers $\tilde{N}_0,\tilde{f}_0^{\mathfrak{n},\lambda},\tilde{k}_0^{\mathfrak{n},\lambda},\tilde{q}_{0}^{\mathfrak{n},\lambda}$ and $\tilde{h}_0^{\mathfrak{n},\lambda}$ are equal to the old ones plus perturbative corrections, which are quadratic in the case of $\tilde{N}_0$ and linear in all other cases \cite{MenaMarugan:2024qnj,MenaMarugan:2025anx}.

The form of the Hamiltonian given in Eq.~\eqref{GIHam} makes it evident that the axial and polar pairs $(Q_{1}^{\mathfrak{n},\lambda},P_{1}^{\mathfrak{n},\lambda})$ and $(Q_{3}^{\mathfrak{n},\lambda},P_{3}^{\mathfrak{n},\lambda})$ describe gauge-invariant degrees of freedom. On the contrary, the configuration variables $Q_{2}^{\mathfrak{n},\lambda}, Q_{4}^{\mathfrak{n},\lambda}, Q_{5}^{\mathfrak{n},\lambda}$ and $Q_{6}^{\mathfrak{n},\lambda}$, whose canonical momenta are perturbative constraints, are not physical. In fact, the gauge-invariant pairs can be chosen such that \cite{Lenzi:2025man,Lenzi:2025pmk}
\begin{align}
\begin{split}
\Theta^{\rm{ax}}&= \frac{1}{32\pi}\sum_{\lambda\in\{\pm\}}\sum_{\mathfrak{n}\in\mathfrak{N}_2}|\tilde{p}_c|\left[(P_{1}^{\mathfrak{n},\lambda})^2 +(Q_{1}^{\mathfrak{n},\lambda})^2\,(\omega_n^2-V^{\rm{ax}}_{l})\right]~,\\
\Theta^{\rm{po}}&=\frac{1}{32\pi}\sum_{\lambda\in\{\pm\}}\sum_{\mathfrak{n}\in\mathfrak{N}_2}|\tilde{p}_c|\left[(P_{3}^{\mathfrak{n},\lambda})^2 +(Q_{3}^{\mathfrak{n},\lambda})^2\,(\omega_n^2-V^{\rm{ax}}_{l})\right]~,
\end{split}
\end{align}
where we have defined the following function of the homogeneous background
\begin{align}\label{Vax}
V^{\rm{ax}}_{l}&=\frac{1}{\tilde{p}_c^2}\left[l(l+1)\left(\tilde{\Omega}^2_b+2\tilde{\Omega}_b\tilde{\Omega}_c\right)-6\tilde{\Omega}_b\tilde{\Omega}_c\right]~,
\end{align}
with $\omega_n=2\pi|n|/L_o$. When evaluated on classical solutions of the unperturbed background, this function turns out to be equal to the Regge-Wheeler potential for gravitational waves in the interior of Kruskal spacetime. The gauge-invariant pairs $(Q_{1}^{\mathfrak{n},\lambda}, P_{1}^{\mathfrak{n},\lambda})$ and $(Q_{3}^{\mathfrak{n},\lambda}, P_{3}^{\mathfrak{n},\lambda})$ have, thus, a clear physical interpretation as gravitational wave modes. For completeness, a proof of this statement is included in the Appendix.

It may seem surprising that the polar Hamiltonian $\Theta^{\rm{po}}$ displays the same potential as the axial one. Indeed, what is most commonly found in the literature is that polar gauge invariants are subject to the so-called Zerilli potential \cite{Zerilli:1970se}. However, this is simply the result of making a particular choice of variables to describe the polar perturbations: The Chandrasekhar transformation \cite{Chandrasekhar:1985kt}, which can be viewed as a background-dependent canonical transformation in the Hamiltonian framework for the full system \cite{Lenzi:2025pmk}, makes it possible to relate the Zerilli variables with our pair $(Q_{3}^{\mathfrak{n},\lambda}, P_{3}^{\mathfrak{n},\lambda})$. Our choice is motivated by the simplicity of the Regge-Wheeler potential, which makes its quantization particularly amenable in LQC.

The Hamiltonian formulation of the system in terms of gauge invariants is an ideal starting point for its quantization. As mentioned in the Introduction, we will follow a hybrid strategy that combines a loop quantization for the background sector of phase space with a Fock representation of the gauge-invariant perturbations. In this case scenario, this means that we adopt the quantum representation explained in Subsec.~\ref{subsec:2.1} for the background variables $(\tilde{\Omega}_b,\tilde{p}_b,\tilde{\Omega}_c,\tilde{p}_c)$, on the separable sector $\mathcal{H}_b\otimes\mathcal{H}_c$ of the Hilbert space in LQC. This allows us to construct operator representations of the background-dependent functions appearing in the Hamiltonian (see Subsec.~\ref{subsec:3c}). On the other hand, a physical criterion is needed to fix a Fock representation for the infinite tower of mode pairs $(Q_{1}^{\mathfrak{n},\lambda}, P_{1}^{\mathfrak{n},\lambda})$ and $(Q_{3}^{\mathfrak{n},\lambda}, P_{3}^{\mathfrak{n},\lambda})$. Since the results in this work do not critically depend on this freedom, we simply assume that a suitable choice of Fock vacuum can be made and adhere to its corresponding representation. For instance, Refs.~\cite{MenaMarugan:2024qnj,MenaMarugan:2025anx} suggest using any vacua belonging to the unique equivalence family that admits a unitarily implementable dynamics, and an ultraviolet diagonalization of the Hamiltonian, in the context of Quantum Field Theory in curved spacetimes of Kantowski-Sachs type \cite{Cortez:2023hux,Cortez:2024soq}. In any case, we respectively call $\mathcal{F}_{\rm{ax}}$ and $\mathcal{F}_{\rm{po}}$ the resulting Fock spaces from the quantization of the fields described by the mode pairs $(Q_{1}^{\mathfrak{n},\lambda}, P_{1}^{\mathfrak{n},\lambda})$ and $(Q_{3}^{\mathfrak{n},\lambda}, P_{3}^{\mathfrak{n},\lambda})$.

\section{\label{sec:3} Quantum Hamiltonian constraint}

The hybrid quantization strategy provides a representation of the zero-mode of the Hamiltonian constraint [multiplying $\tilde{N}_0$ in Eq.~\eqref{GIHam}], on a Hilbert space of the form $\mathcal{H}_b\otimes \mathcal{H}_c\otimes\mathcal{F}_{\rm{pert}}$, where $\mathcal{F}_{\rm{pert}}=\mathcal{F}_{\rm{ax}}\otimes\mathcal{F}_{\rm{po}}$ is the total Fock space of the inhomogeneous perturbations. In principle, at the kinematic level, quantum states would also depend on the perturbative degrees of freedom that are pure gauge. However, we have completely isolated these from the rest of phase space, and identified their canonical momenta as perturbative constraints. It is, then, not hard to convince oneself that physical states in the quantum theory must only contain information about the cosmological background and the (axial and polar) gauge invariants. Thus, in what follows we consider $\mathcal{H}_{\rm{kin}}=\mathcal{H}_b\otimes \mathcal{H}_c\otimes\mathcal{F}_{\rm{pert}}$ as the relevant sector of the kinematic Hilbert space.

Following Dirac's approach for the quantization of constrained systems \cite{dirac2001lectures}, physical states must be annihilated by the (adjoint) action of the constraint operators. According to our comments above, the only one that is non-trivial to impose at the quantum level is the zero-mode of the Hamiltonian constraint. In this section, we study the equations resulting from its imposition on families of states that are of physical interest, namely those for which the effect of the perturbative inhomogeneities on the homogeneous background is small.

\subsection{Ansatz for physical states}\label{subsec:3a}

Let us consider any quantum state in $\mathcal{H}_{\rm{kin}}$\footnote{Strictly speaking, candidates for physical states belong to the algebraic dual of a dense subspace of $\mathcal{H}_{\rm{kin}}$.} with wavefunction of the form $\Xi(\tilde{o}_b,\tilde{o}_c,\mathcal{N}_{\rm{ax}},\mathcal{N}_{\rm{po}})$, where $\tilde{o}_j$ ($j=b,c$) generically denotes quantum numbers of the two background sectors of Hilbert space, and $\mathcal{N}_{\rm{ax}},\mathcal{N}_{\rm{po}}$ are the occupancy numbers on $n-$particle states of the Fock representation chosen for, respectively, axial and polar modes. 

Our goal is to impose the zero-mode of the Hamiltonian constraint on wavefunctions that describe states of physical interest, considering the perturbative nature of the model. In particular, we want to focus on scenarios where the quantum backreaction of the perturbative degrees of freedom on the exact black hole geometry can be kept under control, and even ignored in a suitable limit. For this purpose, we restrict attention to quantum states whose wavefunctions factorize into a background and a perturbative part. We shall refer to this class of states as background–perturbation factorized states, or BP-states. Explicitly, they are of the form
\begin{align}\label{BO}
\begin{split}
&\Xi(\tilde{o}_b,\tilde{o}_c,\mathcal{N}_{\rm{ax}},\mathcal{N}_{\rm{po}})=\Gamma(\tilde{o}_b,\tilde{o}_c)\chi(\tilde{o}_c,\mathcal{N}_{\rm{ax}},\mathcal{N}_{\rm{po}})~,\\ & \quad\quad\quad\quad\quad\quad\quad\quad\rm{or} \\ &\Xi(\tilde{o}_b,\tilde{o}_c,\mathcal{N}_{\rm{ax}},\mathcal{N}_{\rm{po}})=\Gamma(\tilde{o}_b,\tilde{o}_c)\chi(\tilde{o}_b,\mathcal{N}_{\rm{ax}},\mathcal{N}_{\rm{po}})~.
\end{split}
\end{align}

Two remarks are in order regarding the physical properties of our ansatz. First, the (two-dimensional) sector of the black hole phase space that is not factorized, plays a relational role: The variations of the wavefunction components, $\Gamma$ and $\chi$, with respect to this sector provides a notion of quantum evolution. Second, for each fixed value of the relational variable $\tilde{o}_j$, the states allowed by the ansatz are separable with respect to the tensor product of the non-relational background sector of the kinematic Hilbert space and the perturbative Fock space $\mathcal{F}_{\rm{pert}}$. In this sense, at the kinematic level our ansatz precludes entanglement between the non-relational background degrees of freedom and the perturbative modes.

Given the BP-states as in Eq.~\eqref{BO}, the requirement that the quantum backreaction from the perturbations be small involves choosing $\Gamma$ to be close to the wavefunction of a physical state of the unperturbed model. Any such state must be annihilated by the operator representing the Hamiltonian constraint $C_0$ of the homogeneous spacetime. Solutions to this quantum constraint in LQC can be characterized by whether they admit continuous or discrete values of the black hole mass \cite{Zhang:2021wex,ElizagaNavascues:2023gow}. As mentioned in the Introduction, in this work, we focus on the continuous case. In this scenario, physical states are found to be in $\mathcal{H}_c\otimes \mathcal{D}^{\star}_b$, where $\mathcal{D}^{\star}_b$ is the algebraic dual of the dense subspace of $\mathcal{H}_b$ spanned by the discrete basis of states $|\mu_b\rangle$ [see Eq.~\eqref{polymeric}]. Explicitly, their wavefunction is of the following form \cite{ElizagaNavascues:2023gow}:
\begin{align}
\Gamma_0 (\omega_c,\mu_b)=\xi(\omega_c)\breve\psi(\mu_b,\omega_c)~,
\end{align}
where $\xi$ is any function in $L^{2}(\mathbb{R})$, we recall that $\omega_c$ describes the black hole mass in the quantum theory, and $\breve\psi$ is fixed by the constraint equation. In particular, this latter function has a known asymptotic behavior in the regime of large $|\mu_b|$, given by
\begin{align}\label{psiasymp}
\breve\psi(\mu_b,\omega_c)\sim \frac{1}{\mu_b}e^{a_{(+)}\mu_b}\left[A\exp\left(i\sqrt{\frac{L_o^2\omega_c^2}{1+\delta_b^2}}\ln\mu_b\right)+ \bar{A}\exp\left(-i\sqrt{\frac{L_o^2\omega_c^2}{1+\delta_b^2}}\ln\mu_b\right)\right]~,
\end{align}
where $A$ is a complex constant, $\delta_b$ is a function of $\omega_c$, and
\begin{align}
a_{(+)}=\frac{1}{2}\ln\left(\sqrt{1+\delta_b^2}+|\delta_b|\right)>0~.
\end{align}

BP-states with small backreaction must have $\Gamma$ be close to $\Gamma_0$. Now, the dependence of $\Gamma_0$ on the $b-$sector of the black hole phase space clearly shows a divergent behavior in the Hilbert space $\mathcal{H}_b$, due to the global exponential factor in Eq.~\eqref{psiasymp}. On the other hand, under mild conditions on $\delta_b$, its square integrability with respect to $\omega_c \in \mathbb{R}$ is guaranteed. For this reason, it is most convenient to restrict all attention to ans\"atze corresponding to the second option in Eq.~\eqref{BO}, and regard the $b-$sector of phase space as the relational one for evolution. This allows, in particular, to define a notion of dynamics for all the statistical moments of quantum observables of the $c-$sector, using the Lebesgue probability measure for the black hole mass.

\subsection{Choice of relational time}\label{subsec:3b}

Based on the discussion above, we focus on wavefunctions of the form $\Gamma(\tilde{o}_b,\tilde{o}_c)\chi(\tilde{o}_b,\mathcal{N}_{\rm{ax}},\mathcal{N}_{\rm{po}})$, as a promising ansatz in the quest for physical states. The next step is to make an appropriate choice for the relational quantum number $\tilde{o}_b$. This is important because the factorization of the wavefunction between the background $c-$sector and the perturbations need not simultaneously hold for two different choices of $\tilde{o}_b$, if these are associated with non-commuting operators. 

We can illustrate this phenomenon in a simple manner as follows. Take three particles in standard quantum mechanics: One may consider a wavefunction that factorizes its dependence on two of the particles in position space, but this factorization is, in general,  lost in momentum space simply because the Fourier transform of a product is not the product of Fourier transforms. 

In the unperturbed system, there are two operators of the $b$-sector whose spectral properties are well under control: The (rescaled) triad operator $\hat{p}_b/\delta_b$ and $\hat{\Omega}_b$ (or functions of only one of them). Both of their respective quantum numbers $\mu_b$ and $\omega_b$ would, in principle, be acceptable choices of relational variables. In this scenario, we have a generalized Fourier transform [see Eq.~\eqref{gfourier}] relating both representation spaces, but the issue with a simultaneous factorization persists. We, thus, find ourselves at a crossroads: It is either possible to explore the physics emerging from each choice (and compare them), or invoke a stronger criterion that singles one of them out.

In order to find a way out of this dilemma, it is most useful to revisit the purpose of having a sector of phase space play a relational role: To allow for a notion of dynamics in quantum gravity (a fully constrained Hamiltonian system)~\cite{Rovelli:1995fv}. In this context, it is most desirable that the relational variable used is continuous and displays a classical evolution that is monotonic in the Schwarzschild interior time. This is especially convenient if one wishes to eventually make contact with the classical theory, at least in coordinate systems where it is analytically solvable. For instance, in homogeneous and isotropic LQC, one often chooses a massless scalar field as the continuous relational variable fulfilling this property~\cite{Ashtekar:2006wn,Gomar:2015oea}.

Remarkably, between the operators $\hat{p}_b/\delta_b$ and $\hat{\Omega}_b$, only the latter possesses a continuous spectrum. Moreover, in the $b-$sector of the classical black hole interior,  only $\Omega_b$ turns out to display a monotonic behavior, at least for time coordinates $\tau$ that share the same arrows of time as the Schwarzschild lapse $N_0=\pm L_o^2/p_b^2$. Let us show this in more detail. The dynamical equations for $p_b$ and $\Omega_b$ generated by the classical Kantowski-Sachs Hamiltonian $N_0\,C_0$ are given by:
\begin{align}
\dot{p}_b=N_0\left(\Omega_b+\Omega_c\right)p_b~,\quad \dot{\Omega}_b=-N_0\frac{p_b^2}{L_o^2}~,
\end{align}
where we recall that $\Omega_c$ is a constant of motion. Additionally, the condition that $C_0$ must vanish on classical solutions for all $N_0$ restricts $\Omega_b$ to be related with $p_b$ via
\begin{align}\label{branch}
\Omega_b=-\Omega_c\pm \sqrt{\Omega_c^2-p_b^2/L_o^2}~.
\end{align}
These equations immediately show that the classical variable $\Omega_b$ is monotonic in time if $N_0$ never vanishes\footnote{Notice that $p_b/L_o$ must be real in the black hole interior.}. On the contrary, the instant when $p^2_b/L_o^2=\Omega_c^2$ corresponds to either a maximum or a minimum of $p_b$: We clearly have that $\dot{p}_b=0$ there and $|p_b|/L_o$ cannot grow larger than $|\Omega_c|$, else the variable $\Omega_b$ would not be real. This shows that $p_b$ does not evolve monotonically in the black hole interior. Since $\delta_b$ is constrained to be a function of only the black hole mass $\Omega_c$, the variable $p_b/\delta_b$ is also non-monotonic.

Returning to our factorization ansatz for physical states, we are now justified to fix the relational variable of the $b-$sector to be $\tilde{\omega}_b$: The quantum number associated with the LQC operator representing $\tilde{\Omega}_b$ in the perturbed system. It is noteworthy that beyond being continuous and classically monotonic in time, this choice offers additional advantages. Indeed, a look at Eq.~\eqref{Vax} reveals that {\emph {all dependence}} of the perturbative (gauge-invariant) Hamiltonian on the relational sector is precisely through $\tilde{\Omega}_b$. In the Appendix, we show that this property is genuine of this variable, being unachievable by $\tilde{p}_b$ even through perturbative redefinitions of the lapse function $\tilde{N}_0$.

\subsection{Constraint equation}\label{subsec:3c}

As mentioned earlier, the only constraint that is non-trivial to impose at the quantum level is the zero-mode of the Hamiltonian constraint. Recall that it is given by the sum of $\tilde{C}_0$, i.e. the constraint of the unperturbed model, and the two Hamiltonians, $\Theta^{\rm{ax}}$ and $\Theta^{\rm{po}}$, for the gauge-invariant perturbations. Its hybrid quantization is remarkably simple:
\begin{enumerate}
    \item [1)] $\tilde{C}_0$ is promoted to the constraint operator of the unperturbed model in LQC.
    \item[2)] The gauge-invariant pairs $(Q_{i}^{\mathfrak{n},\lambda},P_{i}^{\mathfrak{n},\lambda})$, $i=1,3$, become appropriate linear combinations of annihilation and creation operators on the chosen Fock spaces.
    \item[3)] Powers of the background functions $\tilde{\Omega}_j$, $j=b,c$, appearing in $\Theta^{\rm{ax}}$ and $\Theta^{\rm{po}}$ are directly replaced by their corresponding LQC operators, already defined in the absence of perturbations. All the holonomy corrections to the perturbative Hamiltonians coming from the LQC representation are encapsulated by these operators.
    \item[4)] The background variable $|\tilde{p}_c|$ and its inverse, appearing in $\Theta^{\rm{ax}}$ and $\Theta^{\rm{po}}$, are quantized making use of an LQC representation for $\tilde{p}_c/\delta_c$, given in Eq.~\eqref{polymeric}, and the spectral theorem. Notice that the operator representing $\tilde{p}_c/\delta_c$ never vanishes when acting on $\mathcal{H}_c$, so this procedure is well-defined\footnote{Another possibility to quantize the inverse powers of $\tilde{p}_c/\delta_c$ would be to employ the so-called Thiemann's trick in LQC \cite{THIEMANN1996257,Ashtekar:2011ni}. This choice would incorporate inverse-triad corrections to the perturbative Hamiltonians. Our subsequent analysis is valid for either route of quantization.}. In particular, factors of $|\delta_c|$ and its inverse appear when following it. These are imposed to be functions of the black hole mass $\tilde{\omega}_c$, as dictated by the unperturbed model.
\end{enumerate}
In the last two steps, factor ordering ambiguities arise in products of non-commuting operators of the $c-$sector. In order to keep the discussion as general as possible, we will not explicitly resolve these ambiguities and denote each such product as a single operator incorporating the chosen ordering. All we assume is that the resulting operators are well-defined on $\mathcal{H}_c$\footnote{For instance Refs.~\cite{MenaMarugan:2024qnj,MenaMarugan:2025anx,Lenzi:2025pmk} propose a viable factor-ordering prescription consisting on taking an algebraic symmetrization with respect to powers of $|\tilde{p}_c/\delta_c|$, on products between such powers and $\tilde{\Omega}_c$.}.

Having fixed a loop quantization for the unperturbed model, and a factor ordering for the $c-$sector, all remaining ambiguities in the quantum representation of the Hamiltonian constraint are those of standard Quantum Field Theory. These only affect the perturbative operators acting on Fock space, and they do not play a prominent role in the following discussion.

The procedure above defines a hybrid operator representing the zero-mode of the Hamiltonian constraint. Our target states, with wavefunctions of the form $\Gamma(\tilde{\omega}_b,\tilde{o}_c)\chi(\tilde{\omega}_b,\mathcal{N}_{\rm{ax}},\mathcal{N}_{\rm{po}})$, are physical only if they are annihilated by this operator, a condition we now impose. To this end, and without loss of generality, we fix $\tilde{o}_c$\,=\,$\tilde{\omega}_c$, i.e. the quantum number associated with the operator representing the variable $\tilde{\Omega}_c$. This quantity physically describes the black hole mass in the quantum theory. The constraint equation on our states then reads as follows:
\begin{align}\label{qconstr}
\varepsilon(\tilde{\omega}_b,\tilde{\omega}_c)\chi+\frac{\delta_b^2(\tilde{\omega_c})}{2L_o}\left[\hat{\mathfrak{p}}_b^2(\Gamma)\chi-\hat{\mathfrak{p}}_b^2(\Gamma\chi)\right]+\hat{\Theta}^{\rm{ax}}(\Gamma\chi)+\hat{\Theta}^{\rm{po}}(\Gamma\chi)=0~,
\end{align}
where $\hat{\mathfrak{p}}_j$ is the operator representing $\tilde{p}_j/\delta_j$, $j=b,c$, and $\delta_b=\delta_b(\tilde{\omega}_c)$ is the relation constraining the regularization parameter $\delta_b$ to be a function of the black hole mass in LQC. In this expression, round parentheses next to an operator indicate its range of action. For example, we have:
\begin{align}
\hat{\mathfrak{p}}_b^2(\Gamma)=\frac{\gamma^2}{4}\sum_{\tilde{\mu}_b}\tilde{\mu}_b^2\,\breve{\Gamma}(\tilde{\mu}_b,\tilde{\omega}_c)\bar{e}^{\epsilon_b}_{\tilde\omega_b}(\tilde\mu_b),
\end{align}
where $\tilde{\mu}_b$ are the eigenvalues of the operator $\hat{\mathfrak{p}}_b$ and $\breve{\Gamma}$ is the generalized Fourier transform of $\Gamma$ with respect to the $b-$sector [see Eqs.~\eqref{polymeric} and \eqref{gfourier} and the discussion around them].

In the constraint equation, we have defined the following function:
\begin{align}
\varepsilon(\tilde{\omega}_b,\tilde{\omega}_c)=-\frac{L_o}{2}\left[(\tilde{\omega}_b^2+2\tilde{\omega}_b\tilde{\omega_c})\Gamma +\frac{\delta_b^2(\tilde{\omega_c})}{2L_o}\hat{\mathfrak{p}}_b^2(\Gamma)\right],
\end{align}
which is nothing other than the action of the constraint of the unperturbed model on the state of $\mathcal{H}_b\otimes\mathcal{H}_c$ with wavefunction $\Gamma$. This quantity becomes smaller the closer $\Gamma$ is to an exact solution of the unperturbed LQC model. It can, then, be understood as a {\emph{quantum backreaction}} term. 

Perturbation theory relies on the assumption that the inhomogeneities should only affect the background spacetime in a controllably small way. Based on this premise, in the following,  we choose $\Gamma$ so that the backreaction term $\varepsilon$ is, at most, of quadratic perturbative order.

\section{\label{sec:4} Effective Regge-Wheeler equations}

The quantum constraint Eq.~\eqref{qconstr} must be exactly satisfied by any physical state within our ansatz. However, its mathematical complexity makes it extremely difficult (if not impossible) to address the task of finding analytical solutions. Fortunately, keeping the physics of the perturbative system in mind, it is possible to make approximations that significantly simplify the issue.

The first approximation is of the mean-field type, and it goes as follows. Notice that the only operators of the $c-$sector that do not act by multiplication on our wavefunctions are those representing powers of $|\tilde{p}_c|$. These appear in the perturbative contributions $\hat{\Theta}^{\rm{ax}}$ and $\hat{\Theta}^{\rm{po}}$ to the constraint. Let us assume that the state with wavefunction $\Gamma$ is chosen so that these operators do not mediate significant transitions between it and other states in the $c-$sector. If this is the case, no information is lost in the constraint Eq.~\eqref{qconstr} if one takes its inner product with $\Gamma$ in the Hilbert space $\mathcal{H}_c$. The result is:
\begin{align}\label{effconstr}
\varepsilon_{\Gamma}(\tilde{\omega}_b)\chi+\hat{B}_{\Gamma}(\chi)+\langle\hat{\Theta}^{\rm{ax}}\rangle_{\Gamma}(\chi)+\langle\hat{\Theta}^{\rm{po}}\rangle_\Gamma(\chi)=0~,
\end{align}
where $\langle \cdot\rangle_{\Gamma}$ denotes the expectation value $\bra{\Gamma} \cdot \ket{\Gamma}$ with respect to the inner product of $\mathcal{H}_c$. We have defined the backreaction quantity
\begin{align}
\varepsilon_{\Gamma}(\tilde{\omega}_b)=\int_{-\infty}^{\infty}d\tilde{\omega}_c\,\bar{\Gamma}(\tilde{\omega}_c,\tilde{\omega}_b)\varepsilon(\tilde{\omega}_b,\tilde{\omega}_c)~.
\end{align}
Finally, the operator $\hat{B}_{\Gamma}$ is such that, if we denote $\Phi(\tilde{\omega}_b,\tilde{\omega}_c,\mathcal{N}_{\rm{ax}},\mathcal{N}_{\rm{po}})=\Gamma(\tilde{\omega}_b,\tilde{\omega}_c)\chi(\tilde{\omega}_b,\mathcal{N}_{\rm{ax}},\mathcal{N}_{\rm{po}})$, then
\begin{align}\label{Bop}
\hat{B}_{\Gamma}(\chi) =\frac{\gamma^2}{8L_o}\int_{-\infty}^{\infty}d\tilde{\omega}_c\,\bar\Gamma(\tilde{\omega}_b,\tilde{\omega}_c)\delta_b^2(\tilde{\omega}_c)\sum_{\tilde{\mu}_b}\tilde{\mu}_b^2\bigg[&\breve{\Gamma}(\tilde{\mu}_b,\tilde{\omega}_c)\chi(\tilde{\omega}_b,\mathcal{N}_{\rm{ax}},\mathcal{N}_{\rm{po}})\nonumber \\&-\breve{\Phi}(\tilde{\mu}_b,\tilde{\omega}_c,\mathcal{N}_{\rm{ax}},\mathcal{N}_{\rm{po}})\bigg]\bar{e}^{\epsilon_b}_{\tilde\omega_b}(\tilde\mu_b)~. 
\end{align}

It is noteworthy that, since the generalized Fourier transform between $\mathcal{H}_b$ and $L^2(\mathbb{R})$ is a linear operation, it directly follows that $\hat{B}_{\Gamma}$ is linear too: For any two partial wavefunctions $\chi_{1},\chi_2$, and complex numbers $a_1,a_2$, it holds that $\hat{B}_{\Gamma}(a_1\chi_1+a_2\chi_2)=a_1\hat{B}_{\Gamma}(\chi_1)+a_2\hat{B}_{\Gamma}(\chi_2)$.

After the mean-field approximation, the quantum constraint equation {\emph{effectively}} reduces to imposing that the state with wavefunction $\chi(\tilde{\omega}_b,\mathcal{N}_{\rm{ax}},\mathcal{N}_{\rm{po}})$ be annihilated by the following linear operator:
\begin{align}
\hat{\mathcal{C}}_{\rm{eff}}=\varepsilon_{\Gamma}(\tilde{\omega}_b)+\hat{B}_{\Gamma}+\langle\hat{\Theta}^{\rm{ax}}\rangle_{\Gamma}+\langle\hat{\Theta}^{\rm{po}}\rangle_\Gamma~.
\end{align}
In particular, we find that
\begin{align}
\langle\hat{\Theta}^{\rm{ax}}\rangle_{\Gamma}=\frac{1}{32\pi}\sum_{\lambda\in\{\pm\}}\sum_{\mathfrak{n}\in\mathfrak{N}_2}\left[\langle\widehat{|\tilde{p}_c|}\rangle_\Gamma(\hat{P}_{1}^{\mathfrak{n},\lambda})^2 +(\hat{Q}_{1}^{\mathfrak{n},\lambda})^2\left(\omega_n^2\langle\widehat{|\tilde{p}_c|}\rangle_\Gamma-\langle\hat{\mathcal{V}}^{\rm{ax}}_{l}\rangle_{\Gamma}\right)\right]~,
\end{align}
and similarly for $\langle\hat{\Theta}^{\rm{po}}\rangle_\Gamma$, where $\hat{P}_{1}^{\mathfrak{n},\lambda}$ and $\hat{Q}_{1}^{\mathfrak{n},\lambda}$ are appropriate products of annihilation and creation operators on the Fock space $\mathcal{F}_{\rm{ax}}$, and
\begin{align}
\langle\hat{\mathcal{V}}^{\rm{ax}}_{l}\rangle_{\Gamma}=l(l+1)\,\tilde{\omega}_b\,\left\langle 2|\widehat{\tilde{p}_c|^{-1}\tilde{\Omega}_c}+\tilde{\omega}_b\,\widehat{|\tilde{p}_c|}^{-1}\right\rangle_\Gamma -6\,\tilde{\omega}_b\left\langle|\widehat{\tilde{p}_c|^{-1}\tilde{\Omega}_c}\right\rangle_\Gamma~.
\end{align}
Let us note that this partial expectation value is a function only of the relational variable $\tilde{\omega}_b$.

The operator $\hat{\mathcal{C}}_{\rm{eff}}$ can be interpreted as an effective constraint for the perturbations, at least as long as it is of quadratic perturbative order. By assumption, $\varepsilon_{\Gamma},\hat{\Theta}^{\rm{ax}}$, and $\hat{\Theta}^{\rm{po}}$ are of this order. The only quantity that may not display this property is the operator $\hat{B}_{\Gamma}$. However, based on Eq.~\eqref{Bop}, we reasonably expect that its action on $\chi$ be small if the states with wavefunctions $\Gamma$ and $\Gamma\chi$ are both peaked on the same background geometry, semiclassically described by some metric of the form \eqref{backgroundmetric}. This should be the case if the quantum backreaction from the inhomogeneities on the background can be treated perturbatively, which is our guiding assumption.

In any case, and regardless of the ``smallness" of $\hat{B}_\Gamma$, the effective constraint $\hat{\mathcal{C}}_{\rm{eff}}$ only depends on the Fock operators representing the gauge invariants via $\langle\hat{\Theta}^{\rm{ax}}\rangle_\Gamma$ and $\langle\hat{\Theta}^{\rm{po}}\rangle_\Gamma$. This dependence is quadratic. Moreover, the partial expectation values only depend on the $b-$sector of the background through the relational variable $\tilde{\omega}_b$. Each of them can, thus, be viewed as an explicitly $\tilde{\omega}_b$-dependent Fock Hamiltonian of a linear field theory. In this framework, each such Hamiltonian generates an effective dynamics for the perturbative gauge-invariants in some time $T$ (which would correspond to the choice $\tilde{N}_0=1$ in the classical theory). 

Explicitly, the dynamics generated by the effective Hamiltonians are ruled by the following Heisenberg equations, for $i=1,3$:
\begin{align}
\frac{d\hat{Q}_{i}^{\mathfrak{n},\lambda}}{dT}=\langle\widehat{|\tilde{p}_c|}\rangle_\Gamma\hat{P}_{i}^{\mathfrak{n},\lambda},\qquad \frac{d\hat{P}_{i}^{\mathfrak{n},\lambda}}{dT}=-\left(\omega_n^2\langle\widehat{|\tilde{p}_c|}\rangle_\Gamma-\langle\hat{\mathcal{V}}^{\rm{ax}}_{l}\rangle_{\Gamma}\right)\hat{Q}_{i}^{\mathfrak{n},\lambda}~.
\end{align}
The corresponding evolution also has a direct counterpart in the form of effective Hamilton's equations for the classical canonical pairs $(Q_i^{\mathfrak{n},\lambda},P_i^{\mathfrak{n},\lambda})$.

It is possible to make contact between the effective dynamics we just derived and the well-known classical equations for the gauge-invariant perturbations. Let us introduce the following change of time:
\begin{align}\label{dressedtime}
\langle\widehat{|\tilde{p}_c|}\rangle_\Gamma\,dT=d\tau_{\star}~.
\end{align}
Note that it depends on the partial background wavefunction $\Gamma$, even if we omit it in the notation. In this new time, the effective equations are equivalent to
\begin{align}\label{effectivedyn}
\frac{d^2 Q_{i}^{\mathfrak{n},\lambda}}{d\tau_{\star}^2}=-\left(\omega_n^2-\frac{\langle\hat{\mathcal{V}}^{\rm{ax}}_{l}\rangle_{\Gamma}}{\langle\widehat{|\tilde{p}_c|}\rangle_\Gamma}\right)Q_{i}^{\mathfrak{n},\lambda},\quad i=1,3~,
\end{align}
which would be nothing but the Regge-Wheeler equations if the expectation values on $\Gamma$ exactly described the classical unperturbed geometry of the black hole interior. Indeed, note that the classical counterpart of $\hat{\mathcal{V}}^{\rm{ax}}_l$ is nothing but $|\tilde{p}_c|$ times the Regge-Wheeler potential $V_{l}^{\rm{ax}}$.

We conclude that Eq.~\eqref{effectivedyn} can be interpreted as the effective Regge-Wheeler equations for gauge invariants emerging from a loop quantum black hole, with $\tau_{\star}$ being the corresponding tortoise coordinate\footnote{That $\tau_\star$ indeed describes the tortoise coordinate in the classical theory can be directly seen as follows. The $\tau-x$ part of the background metric~\eqref{backgroundmetric} is conformally related to $-N^2_0p^2_c\,d\tau^2+dx^2$, with $p_b^2/(L_o^2|p_c|)$ being the conformal factor. Recalling that $T$ corresponds to $N_0=1$, the classical counterpart of Eq.~\eqref{dressedtime} then yields $-d\tau_{\star}^2+dx^2$ for the conformal metric. This is the standard identification of the tortoise coordinate in Schwarzschild spacetime.}. The expectation values appearing in these equations encode the LQC modifications to the evolution of the gravitational wave modes in the interior black hole region.

\section{\label{sec:5} Conclusions}

We have derived an effective set of equations ruling the dynamics of gravitational perturbations in the interior of a loop quantum black hole with spherical symmetry. As differential equations, they share the same structure as their classical counterpart---the Regge-Wheeler equations---except that the potential contains modifications due to the quantum nature of the black hole geometry. These modifications appear in the form of expectation values of this geometry.

The effective dynamics obtained are valid for quantum states of the black hole that display a controllably small backreaction from the perturbations on the background geometry. This type of states is the most natural to describe any scenario where gravitational waves of small amplitude propagate around a black hole spacetime. A remarkable example of such a scenario would be the ringdown phase of a black hole merger. In this context, the effective equations here derived set the foundations for, ultimately, investigating LQG phenomenology on gravitational wave signals. 

Our study is based on the recent works \cite{MenaMarugan:2024qnj,MenaMarugan:2025anx,Lenzi:2025pmk}, where the authors develop a Hamiltonian formulation and hybrid quantization of the perturbed black hole system using gauge invariants. The hybrid quantization consists of combining a Fock representation for the perturbative fields, which are purely inhomogeneous, and a quantum gravity-inspired quantization for the homogeneous background. We have focused on LQC for the latter, given its recent success in the quantization of the unperturbed model and the attractiveness of effective models based on it \cite{Ashtekar:2018cay,Zhang:2021wex,ElizagaNavascues:2023gow}.

The main result of this article is a consequence of imposing, at the quantum level, the zero-mode of the Hamiltonian constraint of the perturbed system. In order to do this in a physically meaningful way, we have restricted all attention to BP-states: The allowed wavefunctions display a separate dependence on a sector of the background phase space and the gauge-invariant perturbations. The two parts of the wavefunction are connected through the remaining background sector, which acts in a relational manner and allows to define a notion of evolution. Remarkably, this sector has a geometrical interpretation at the classical level, unlike most studies in LQC that use matter as a relational clock \cite{Ashtekar:2006wn,Gomar:2015oea,MenaMarugan:2024qnj,Zhang:2021wex}.

We have resolved the ambiguity in the choice of relational sector of phase space using the properties of physical states of the unperturbed system that allow for a continuous black hole mass. In this case, inner-product divergences of the background sector that classically characterizes the black hole horizon (called $b-$sector here) motivate giving it the relational role. The other ($c-$)sector describes the black hole mass, and physical states are square integrable with respect to it. The complementary analysis of the case where this mass can take discrete values is left for future work.

There are two variables of the $b-$sector that can be used as a relational clock and, at the same time, admit a well-defined representation in LQC. Because of its role in the quantum states, we have chosen the only one that is continuous and displays a monotonic evolution classically. Remarkably, after applying a mean-field approximation on the constraint equation with respect to the $c-$sector, the resulting (effective) constraint only depends on the perturbations and the relational variable. This allows us to view this constraint as a generator of the effective dynamics for the gauge invariants.

A natural continuation of this work would be to analyze in detail the effective equations derived. For this purpose, it proves necessary to relate the relational variable $\Omega_b$ with the time $T$ in which the effective constraint generates evolution. A possibility to do this is to adhere to an effective LQC geometry, e.g. the so-called Ashtekar-Olmedo-Singh model \cite{Ashtekar:2018cay}. In this spacetime model, which describes a black-to-white hole transition, all functions of the background phase space have an analytic evolution. The study of the effective Regge-Wheeler equations in this context would then be feasible if one could insert this evolution in the expectation values that modify the potential term.

Future understanding of the effective evolution will pave the road to extracting physical predictions of hybrid LQC on gravitational waves in the exterior black hole region. Remarkably, Refs.~\cite{Lenzi:2025pmk,Lenzi:2025man} have already taken the first steps to the extension of the canonical model to the exterior. The procedure is based on the realization that the complex canonical transformation $(b,p_b)\rightarrow (-ib,ip_b)$ effectively interchanges the roles of the space and time in the background metric \eqref{backgroundmetric}. The result is a spherically symmetric and static spacetime, that can be interpreted as describing the black hole exterior. Originally, this canonical transformation was used by the Asthekar-Olmedo-Singh model \cite{Ashtekar:2018cay} in order to extend the effective dynamics of the interior model to the exterior, in the absence of perturbations. In the perturbed model, the effective equations derived for the gravitational wave modes display all their dependence on the $b-$sector of the background via a relational variable $\tilde{\omega}_b$ that classically corresponds to the product $\tilde{b}\,\tilde{p}_b/(\gamma L_o)$. This phase space function is invariant under the aforementioned complex canonical transformation, up to terms that are at most quadratic in perturbations. Furthermore, for physically sound choices of quantum states, each background expectation value appearing in the effective equations should (at least locally) admit a  parameterization of the form $\tilde{\omega}_b=\tilde{\omega}_b(\tau_{\star})$, where $\tau_{\star}$ is the effective time variable introduced in Eq.~\eqref{dressedtime}. Since the classical counterpart of this time simply corresponds to the tortoise coordinate in the interior, the invariance of $\tilde{b}\,\tilde{p}_b/(\gamma L_o)$ naturally motivates identifying $\tau_{\star}$ with the tortoise radius $r_{\star}$ in the exterior region, at the level of the effective equations. The main difficulties for extending in this way the effective equations to the exterior black hole region would, then, be rooted at determining the specific parameterization $\tilde{\omega}_b=\tilde{\omega}_b(\tau_{\star})$, either from the full quantum theory or from the interior dynamics of an effective model (with suitable motivation).

Extending the effective Regge-Wheeler equations derived here to the exterior region, along the lines sketched above, would allow one to study the effects of hybrid LQC on the quasinormal mode spectrum of gravitational waves and its isospectrality. Some of these issues have already been addressed in literature using test fields propagating on an effective metric, ``dressed" with LQC corrections \cite{Daghigh:2020fmw,del-Corral:2022kbk}. As it has been proven in primordial cosmology \cite{ElizagaNavascues:2017avq,MenaMarugan:2024zcv}, the effective dynamics derived from the hybrid model could very well lead to different results.

Finally, it would be most desirable to extract a similar set of effective equations when the black hole states display discrete values for their mass (a possibility allowed in LQC \cite{Zhang:2021wex}). This would allow, by comparison, to analyze the consequences (if any) of such discreteness in the effective propagation of gravitational waves in the interior.

\acknowledgments
The authors are very grateful to G.A. Mena Marugán and A. Mínguez-Sánchez for enlightening discussions throughout. We acknowledge partial support by Project No. PID2023-149018NB-C41 from Spain. A.T.-C. acknowledges
support from the PIPF-2024 fellowship from Comunidad Autónoma de Madrid. The reference number is PIPF-2024/TEC-34465.

\appendix
\section{Some aspects of the perturbative Hamiltonian}\label{Appendix}

We first show that the function on phase space $V_l^{\rm{ax}}$, given in Eq.~\eqref{Vax} is nothing but the Regge-Wheeler potential in the classical linearized theory. There, notice that we can identify the set of perturbatively corrected variables $(\tilde{N}_0,\tilde{c},\tilde{p}_c,\tilde{b},\tilde{p}_b)$ with their untilded counterpart.

Taking the GR solution for the interior black hole region, a look at Eq.~\ref{backgroundmetric} reveals that $|p_c|=r^2$, where $r\in (0,2M)$ is the Schwarzschild time coordinate in the interior. It then follows that, in Schwarzschild coordinates, $p_b^2 = -r^2 f(r) L_o^2$ and $N_0= \pm[r^2 f(r)]^{-1}$, with $f(r) = 1 - 2M/r$ and $M$ denoting the mass of the black hole.

The identifications above allow one to find the classical values of $\Omega_b$ and $\Omega_c$, using the equations of motion. On the one hand, Hamilton's equation for $p_c$ reads as follows:
\begin{equation}
\dot{p}_c = \{p_c,H_{{\rm{KS}}}[N_0]\} = 2N_0 \Omega_b p_c~,
\end{equation}
where, in Schwarzschild coordinates, we have that $\dot p_c =2r\,{\rm{sign}}(p_c)$. Solving for $\Omega_b$ we obtain: $\Omega_b = \pm r f(r)$ for $N_0= \pm[r^2 f(r)]^{-1}$. On the other hand, demanding that the Hamiltonian constraint $C_0$ be zero on classical solutions amounts to the following relation:
\begin{equation}
\frac{p_b^2}{L_o^2}=-\Omega_b^2-2\Omega_b\Omega_c~.
\end{equation}
This is only consistent with the Schwarzschild values of $p_b^2/L_o^2$ and $\Omega_b$ if $\Omega_c=\pm M$ for $\Omega_b = \pm r f(r)$.

Direct substitution in Eq.~\eqref{Vax} of the Schwarzschild values found for $\Omega_b$ and $\Omega_c$ finally yields
\begin{equation}
V^{\rm{ax}}_{l}= f(r) \left[ \frac{l (l+1)}{r^2} - \frac{6M}{r^3}\right]~.
\end{equation}
on classical solutions. This is the Regge-Wheeler potential for gravitational waves (spin-$2$ fields) in Schwarzschild spacetime.

Next, let us show that the dependence of the perturbative Hamiltonian on the relational $b-$sector cannot be expressed exclusively in terms of $\tilde{p}_b$, under any acceptable redefinition of the lapse. 

The total Hamiltonian of the perturbed system (truncated at quadratic perturbative order) is invariant under the following transformation
\begin{align}
V^{\rm{ax}}_l\rightarrow V^{\rm{ax}}_l+F_{l}\,\tilde{C}_0~,\qquad \tilde{N}_0\rightarrow \tilde{N}_0+\frac{1}{32\pi}\sum_{\lambda\in\{\pm\}}\sum_{\mathfrak{n}\in\mathfrak{N}_2}|\tilde{p}_c|F_{l}\left[(Q_{1}^{\mathfrak{n},\lambda})^2 +(Q_{3}^{\mathfrak{n},\lambda})^2\right]~,
\end{align}
where $F_{l}$ is any function on the background sector of phase space that is well-defined in the classical theory.
This symmetry is an off-shell counterpart of the classical freedom to use $\tilde{C}_0=0$ in the perturbative Hamiltonian, when it is evaluated on the constrained surface of phase space (and truncated at quadratic order in perturbations).

We claim that there is no acceptable function $F_{l}$ such that a transformation like the above can transmute the potential $V_l^{\rm{ax}}$ into a function of only $\tilde{p}_b$ and the $c-$sector. Let us assume that this was possible. Then, $F_{l}$ should be such that
\begin{align}
F_{l}\,\tilde{C}_0-\frac{1}{\tilde{p}_c^2}\left[l(l+1)\left(\frac{2}{L_o}\tilde{C}_0+\frac{\tilde{p}_b^2}{L_o^2}\right)+6\tilde{\Omega}_b\tilde{\Omega}_c\right]=G_l(\tilde{p}_b,\tilde{p}_c,\tilde{\Omega}_c)~,
\end{align}
where $G_l$ is an arbitrary function of its argument. Because of the last term in the square brackets, this equality cannot hold unless $F_l$ is a function that diverges when $\tilde{C}_0=0$. This would lead to an ill-defined lapse in the classical theory, something that contradicts our hypothesis.

\bibliography{library}
\end{document}